\documentstyle[prl,aps,preprint]{revtex}
\tightenlines
\begin{document}
\draft
\preprint{hep-ph/9708202}
\title{Proton-Proton Spin Correlations at Charm Threshold\\
               and Quarkonium Bound to Nuclei
}
\author{Guy F. de T\'eramond and  Randall Espinoza}
\address{Escuela de F\'{\i}sica, Universidad de Costa Rica, San Jos\'e, Costa
Rica}
\author{Manuel Ortega-Rodr\'{\i}guez\cite{manuel}}
\address{Department of Applied Physics, Stanford University, Stanford, 
California 94309}
\date{\today}   
\maketitle
\begin{abstract}

The anomalous behavior of the spin-spin correlation at large momentum
transfer in $p\,p$ elastic scattering is described in terms of the
degrees of freedom associated with the onset of the charm threshold. A
non-perturbative analysis based on the symmetries of QCD is used to
extract the relevant dynamics of the charmonium-proton interaction. The
enhancement to $p\,p$ amplitudes and their phase follow from analyticity
and unitarity, giving a plausible explanation of the spin anomaly. The
interaction between $c \bar c$ and light quarks in nuclei may form a
distinct kind of nuclear matter, nuclear bound quarkonium. 
\end{abstract}
\bigskip
\pacs{PACS numbers: +12.38.-t, 13.85.Dz, 13.88.+e, 12.38.Lg, 14.40.Gx, 
24.85.+p, 21.30.-x}
\narrowtext
The fundamental description of the strong forces in terms of structureless 
constituents which carry the symmetries within the hadrons, and their 
interaction in terms of gluonic fields, is the basis of quantum chromodynamics, 
QCD. Despite considerable success in the application of QCD in the 
perturbative regime (PQCD)~\cite{scaling}, there are important discrepancies
with measurements related to the production and decay of heavy quark 
systems~\cite{brod1} and the study of spin effects at high 
energy~\cite{Krisch}. A notorious example is the unexpected result observed 
in the polarization correlation $A_{NN}$ found in proton-proton collisions 
at large energy and momentum transfer~\cite{spin}. At $\sqrt{s}$ near 5~GeV 
and $\theta_{\rm c.m.} = 90 ^\circ$ the rate of protons scattering with 
incident spins parallel and normal to the scattering plane, relative to the 
protons scattering with antiparallel spins, rises dramatically with $A_{NN}$
reaching a value of 60~\% in a region where scaling behavior is expected. 

\medskip
A simple explanation for this anomaly has its origin in the observation
that at $\sqrt{s} = 5\rm \ GeV$ new degrees of freedom associated with
the onset of a heavy quark-antiquark threshold could invalidate the
scale-invariant nature of the basic constituent interaction at high
energies by the interaction of light and heavy quarks at low relative
velocity~\cite{BdT}. Other models had also been advanced to explain the
spin anomaly, but they imply a radical departure from the PQCD framework
or the Standard Model~\cite{GdT}, introducing extraneous dynamical
effects. The phenomenological analysis of Ref.~\cite{BdT}, based on
$J=L=S=1$ broad resonance structures at the strange and charm
thresholds interfering with PQCD quark-interchange amplitudes,
successfully describes the anomalous $p_{\rm lab}$ dependence of the spin
correlation but lacks of a clear dynamical content since the parameters of the
model are chosen to fit the data. 

\medskip
It is one of the main objectives of the present paper to show that the
enhancement of the proton-proton elastic amplitudes close to the charm
threshold is obtained from precise knowledge of the dynamics of heavy
quarkonium with light hadrons. The spin correlation close to the
threshold is determined from unitarity and analyticity without
introducing arbitrary parameters and is largely predictable from basic
principles.

\medskip
The study of the interaction of quarkonia with light hadrons is of
particular relevance to a better understanding of the dynamics of gluons 
from a new perspective. Since different quark flavors
are involved in the quarkonium-nucleon interaction, there is no quark
exchange to first order in elastic scattering~\cite{BSdT} and multiple
gluon exchange is the dominant contribution~\cite{brod-mill}.
Remarkably, this system involves basic QCD degrees of freedom at the
nuclear level, in contrast with the usual description of the dynamics of
the nuclear forces with phenomenological lagrangians in terms of meson
and nucleon degrees of freedom. 

\medskip
Due to the small size of charmonium at hadronic scale, its interaction
with external gluon fields from interacting hadrons is expressed as a
QCD multipole expansion~\cite{peskin} with higher order terms suppressed
by powers of the quarkonium radius over the gluon wavelength, resulting
in a long distance effective theory that has been applied successfully at
the leading order in many processes involving low energy
gluons~\cite{Yan}. The leading term  in the expansion is the bilinear
chromo-electric term, which is in fact the leading local gauge-invariant 
operator in an operator product expansion (OPE), whose matrix elements 
between hadronic states are evaluated using the trace anomaly~\cite{voloshin,mls}. 
This is a notorious example of a calculation where QCD matrix elements, 
otherwise untractable by perturbation theory, are determined by the symmetries 
of the theory. 

\medskip
Using non-relativistic normalization of states and neglecting the
charmonium recoil, the charmonium-nucleon amplitude can be written as a
product of two matrix elements, one depending on the internal charmonium
degrees of freedom and the other containing the external gluon 
fields~\cite{peskin}. After some color algebra the result is 
$$
f = -\frac{m_{\rm red}}{2\pi}\frac{2\pi}{3}\alpha_s 
\langle{\varphi}|r^i\frac{1}{H^{(8)} + \epsilon} r^j |\varphi\rangle
\langle {K_2}| E^a_i E^a_j | {K_1}\rangle,
$$
where $(H^{(8)} + \epsilon)^{-1}$ propagates the $c\bar c$ system in the 
octet state, $m_{\rm red}$ is the reduced mass of the proton-charmonium 
system, $\bf K$ its relative momentum and $\varphi$ is the internal 
charmonium wave function. Standard evaluation of the charmonium matrix 
element gives
$$
f= -\frac{m_{\rm red}}{2\pi} d_2 a_0^3 
\big\langle\frac{1}{2}{\bf E}_a {\bf \cdot} {\bf E}_{a}\big\rangle,
$$
where $d_2 = {28\pi/ 27}$ is the Wilson coefficient for a $1s$ charmonium
state with Bohr radius $a_0$.

\medskip
The field operator ${\bf E}^2$ is expressed as a linear combination of
the gluonic contribution to the energy-momentum tensor $T^{\mu\nu}_{\rm
gluonic}$ and the trace of the total energy-momentum tensor, $T^\mu_\mu$,
through the scale anomaly~\cite{voloshin,mls} renormalized at the
charmonium scale $\Lambda_Q \sim 1/a_0$. At zero momentum transfer the
matrix elements of $T^{\mu\nu}_{\rm gluonic}$ and $T^\mu_\mu$ are
expressed in terms of the gluon-momentum fraction in the nucleon, $V_2$,
and the nucleon mass, $M$, respectively. In this case the value of $f$
equals the scattering length, $a_s$, and is given by
$$
f = -\frac{m_{\rm red}}{2\pi} d_2 a_0^3 \frac{M}{2} \bigg (\frac{3}{4}
 V_2(\Lambda_Q)-\frac{\alpha_s(\Lambda_Q)}{\beta(\alpha_s) }\bigg).
$$
Using in the previous equation a value of $V_2=0.5$ from deep-inelastic
scattering measurements, the hydrogenoid charmonium value $a_0 = 1.6
\rm\ GeV^{-1}$, the value of $\alpha_s = 0.6$ evaluated at the scale
$\Lambda_Q$, and approximating the Gell-Mann--Low beta function by its
leading term $-9\alpha_s^2/2\pi$, we find a value of $a_s = -0.2\rm\ fm$
for the charmonium-proton scattering length, which agrees with the
result obtained by Brodsky and Miller~\cite{brod-mill} from
Ref.~\cite{mls}. The sign of the amplitude corresponds to an attractive
interaction. Other results~\cite{kaidalov,Kharzeev,shevchenko} differ
widely because of different normalizations used in the
calculation~\cite{pie}.

\medskip
Scattering at low energies is described in terms of two parameters, the
scattering length $a_s$, and the effective range $r_{\rm e}$.  We use the value
of the scattering length determined above, $a_s = -0.2\rm\ fm$, to
describe the low energy charmonium-nucleon scattering. The value
of the $c\bar c$-$p$ effective range is unknown and can not be obtained from
the previous calculation, valid only at zero momentum transfer. We
expect however a value for $r_{\rm e}$ between $a_0 \simeq 0.3{\rm\ fm}$
and 1~fm, which accounts for the small size of charmonium and the short
range of the QCD van der Waals force. With these values, we compute the
effect on the $p\,p$~elastic amplitudes at the onset of charm threshold
\cite{BdT} and determine the binding energies of charmonium in nuclear
matter \cite{BSdT,Wasson}. Our results do not depend significantly on
the value of $r_{\rm e}$.

\medskip
The unitarity condition relates the imaginary part of an elastic 
scattering amplitude to a sum over all possible intermediate states that
can be reached from the initial state. The enhancement to a given 
partial-wave amplitude $T$ is determined by the analytic properties of the 
transition amplitude in the $\nu = p^2$ complex plane, with $p$ the cms 
momentum, and is given by the dispersion integral~\cite{fsi}
$$
T^{\rm enh}(\nu)= T(\nu) + \frac{1}{\pi}\int_0^\infty
\frac{\,{\text{Im}\,} T(\nu')}{\nu' - \nu -i\varepsilon} d\nu'.
$$
We assume that the inelastic contribution of unitarity to the $p\,p$ elastic
scattering amplitude is dominated near threshold by the creation of a 
$p\,p\,c\bar c$ state with low relative momenta and internal quantum numbers 
$j=1$, $l=0$, which corresponds to the $J = L = S = 1$ quantum numbers of the 
$p\,p$ system.
 The triplet wave has maximal spin correlation $A_{NN} = 1$, and the $p\,p$ 
orbital angular momentum must be odd since the $c\bar c$ has negative 
intrinsic parity. The $p\,p\,c\bar c$ state is produced with partial width 
characterized by the measured branching ratio $B^{pp}_{c\bar c} = 0.0155$. The 
enhancement to the partial wave amplitude $ T_{JJ} = T^{S=1}_{J=L}$ can be 
represented in the $N/D$ form \cite{Chew} in terms of the discontinuity of 
$T(\nu)$
$$
T_{JJ}^{\rm enh}(\nu) = \frac{1}{D(\nu)}\frac{1}{2\pi i} 
\int\frac{D(\nu'){\,\text{disc}\,} T_{JJ}(\nu')}{\nu - \nu'} d\nu',
$$
where the Jost function, $D(\nu)$, is defined by~\cite{fsi}
$$
D(\nu)=\exp\bigg[-\frac{1}{\pi}\int\frac{\delta(\nu')\, d\nu'} 
{\nu-\nu'-i\epsilon}\bigg],
$$
in terms of the proton-proton-charmonium scattering s-wave phase shift 
$\delta(\nu)$ which, by unitarity, is the phase of the $p\,p$ enhanced 
amplitude. We limit our calculation to the zeroth iteration $D=1$ in the 
previous integral, to obtain the determinantal approximation 
$T^{\rm enh} = T/D = R\, T$ with $R = D^{-1}$ the enhancement factor.

\medskip
The $p\,p$ elastic scattering is conveniently described in terms of the usual
 five helicity amplitudes: 
$\phi_1 = \langle{++}|\phi |{++}\rangle$, 
$\phi_2 = \langle{--}|\phi |{++}\rangle$, 
$\phi_3 = \langle{+-}|\phi |{+-}\rangle$, 
$\phi_4 = \langle{-+}|\phi |{+-}\rangle$ and 
$\phi_5 = \langle{++}|\phi |{+-}\rangle$. At large momentum transfer the PQCD
 quark interchange model (QIM) amplitudes \cite{Gunion} for proton-proton 
scattering dominate over quark annihilation and gluon exchange \cite{Blazey}. 
The helicity conserving amplitudes are related by 
$\phi^{\rm QIM}_1 = 2\phi^{\rm QIM }_3=-2\phi^{\rm QIM}_4$ and are the only 
non-vanishing at leading order \cite{relations}. We assume the same form for 
the QIM amplitudes as in Ref.~\cite{BdT} expressed in terms of the standard 
proton dipole form factor, consistent with nominal $s^{-4}$ scaling law and 
angular distribution~\cite{scaling}. We do not consider any phase dependence on the QIM 
amplitudes since
they cancel out in the present calculation. 
 
\medskip
Since the nominal power-law scaling for the scattering 
amplitude is not preserved individually by each partial wave, a modification
of a given partial amplitude effectively breaks the $s^{-4}$ scaling
behavior of the total amplitude at the charm threshold.
The partial wave expansion of the helicity amplitudes is expressed in terms 
of singlet, coupled and uncoupled triplet amplitudes. Retaining the 
enhancement in the $J=L=S=1$ partial wave amplitude we need only to modify 
the $J = 1$ wave in the uncoupled-triplet partial wave 
expansion~\cite{Wong} 
\begin{eqnarray}
\phi_3 &\sim& \sum_{J=\rm odd} (2 J + 1) T_{JJ}\, d^{J}_{11}\nonumber\\
\phi_4 &\sim& -\sum_{J=\rm odd} (2 J + 1) T_{JJ}\, d^{J}_{-11},\nonumber
\end{eqnarray}
with $d^{J}$ the Wigner coefficients. The enhanced helicity amplitudes 
$\phi^{\rm enh}$ expressed in terms of the QIM helicity amplitudes 
$\phi^{\rm QIM}$, the branching ratio $B^{pp}_{c\bar c}$, and the s-wave 
enhancement factor $R(\nu)$ are 
\begin{eqnarray}
\phi_1^{\rm enh} &=& \phi_1^{\rm QIM},\nonumber\\
\phi_3^{\rm enh} &=& \phi_3^{\rm QIM} + 
                   3 B^{pp}_{c\bar c}(R - 1) T_{11} d^1_{11},\nonumber\\
\phi_4^{\rm enh} &=& \phi_4^{\rm QIM} - 
                   3 B^{pp}_{c\bar c} (R - 1) T_{11} d^1_{-11},\nonumber
\end{eqnarray}
where
$T_{11} = \frac{1}{4}\int_{-1}^{+1}(d^{1}_{11}\phi_3^{\rm QIM} - d^{1}_{-11}
\phi_4^{\rm QIM}) dz$, and $\phi_2^{\rm enh}$ and $\phi_5^{\rm enh}$ are zero
 to leading order.
 
\medskip
For the $pp$-$c\bar c$ state the s-wave phase shift is expressed in terms of
 the effective range approximation as
$
k\cot \delta = -1/a + \frac{1}{2}r k^2,
$
where $k = p - p_0$, with $p_0= 2.32 \ \text{GeV}/c$ the cms momentum 
corresponding to the charm threshold. In this approximation the Jost 
function is obtained directly from the phase shift \cite{fsi} giving for the
 enhancement factor $R(k) = (k + i\alpha) / (k + i \beta)$, where $\alpha$ 
and $\beta$ are related to the scattering parameters by 
$a = (\alpha - \beta)/ \alpha \beta$ and $r = 2/(\alpha - \beta)$. For
definiteness we take $a = 2 a_s$, $r \sim r_{\rm e}$, assuming simple
additivity in the interaction of charmonium with the external fields of
the two protons. Comparison of the prediction for the spin correlation 
$A_{NN}$ at $\theta_{\rm c.m.}=90^\circ$ with the available data near 
the charm threshold, is shown in Fig.~\ref{ann} for different values of 
$r_{\rm e}$.

\medskip
Since the QCD van der Waals force is attractive 
\cite{BSdT,mls,kaidalov,Kharzeev,shevchenko}, it could lead to the formation
 of quarkonium bound to nuclei \cite{BSdT,Wasson}. The discovery of such 
state would unveil the purely gluonic component of the strong forces, a 
genuine QCD effect at the nuclear level, which do not involve meson degrees 
of freedom. In fact, in the absence of valence light-quark exchange there is
 no one-meson standard nuclear potential and the contribution of higher-order
 intermediate meson states is negligible~\cite{brod-mill}. Furthermore, there
 is no short-range nuclear repulsion from Pauli blocking. 
In Ref.~\cite{BSdT} an estimate of the QCD nucleon-charmonium van der Waals 
potential was obtained by rescaling the high-energy meson-nucleon interaction
 described by the Pomeron model \cite{DL}, and determining the coupling and 
range of the interaction parametrized by a Yukawa form, 
$V(r) = -\gamma e^{-\mu r}/r$, by adjusting the meson form factor in the 
Pomeron amplitude to describe the size of charmonium. The scattering 
length in the Born approximation, $2 \gamma m_{\rm red}/\mu^2 \simeq
 0.5~\text{fm}$, overestimates the strength of the interaction by almost a 
factor of three compared to the QCD non-perturbative result obtained above.
 
\medskip
Due to the short range of the nuclear forces, additivity is not valid for 
all the nuclei but the very light and the interaction with charmonium 
 depends on the nucleon distribution. Following Wasson \cite{Wasson}, we 
write the charmonium-nucleus potential, $V_{c\bar c\text{-}A}(r)$, at low 
energies as
$$
V_{c\bar c\text{-}A}(r) = \int V_{c\bar c\text{-}N}(r - r')\rho(r') d^3 r',
$$ 
where $\rho(r)$ is the nucleon distribution in the nucleus of nucleon number 
$A$ and $V_{c\bar c\text{-}N}(r)$ the charmonium-nucleon potential. Since the
density in the central core of nuclei is practically constant, falling 
sharply at the edges, and the range of $V_{c\bar c\text{-}N}(r)$ is very 
short compared to the size of nuclei, the above expression for 
$V_{c\bar c\text{-}A}(r)$ is approximated by the nuclear matter result
$$
V_{c\bar c\text{-}A} = \rho_0\int V_{c\bar c\text{-}N}(r) d^3 r = 
\frac{4 \pi \rho_0 a_s}{2 m_{\rm red}} = -11\rm\ MeV,
$$
\nobreak
for $\rho_0 = 0.17\rm\ fm^{-3}$ and $m_{\rm red}$ the $c\bar c$-$N$ 
reduced mass, a result consistent with Ref.~\cite{mls}, and 
comparable with the binding energy of protons and neutrons in nuclear matter. For finite 
nucleus, we describe the nuclear density by the Fermi density function 
$\rho(r) = \rho_0/(1 + e^{(r - c)/b})$ which fits very well the data and 
incorporates the thickness of the nuclear surface~\cite{Uberall}. For the 
light nuclei we have used various forms for $\rho (r)$ found in the 
literature to describe the data~\cite{Uberall}. We have investigated the 
binding of charmonium to various nuclei using different forms for 
$V_{c\bar c\text{-}N}(r)$ which reproduce the same scattering length 
$a_s = -0.2 \rm\ fm$ for various potential ranges. Consistent results were 
obtained for Yukawa, Gaussian, and Bargmann potentials. As an example, we 
give in Table~\ref{tabla} the results obtained for a Gaussian form 
$V(r) = V_0 e^{-(r/R)^2}$, with $a_s = \sqrt{\pi}m_{\rm red} V_0 R^3/2$.

\medskip
We have determined the low energy interaction of heavy quarkonium and
nucleons in a model independent way using the operator product expansion
and the trace anomaly. The value of $-0.2\rm\ fm$ for the 
$c\bar c$-nucleon scattering length corresponds to a total cross section of
about 5~mb at threshold. This value reproduces the anomalous behavior of
the spin correlation $A_{NN}$ observed at $\sqrt{s} = 5\rm\ GeV$ in
$p\,p$ elastic scattering at $\theta_{\rm c.m.} = 90^\circ$ near the
charm threshold, by the enhancement of $p\, p$ amplitudes from the
unitarity condition. The low energy charmonium-nucleon interaction only
depends on the short range gluonic exchange properties of QCD between
color singlet objects and do not involve meson degrees of freedom.
Binding of charmonium was investigated for light, medium and heavy
nuclei using realistic nucleon distributions for various forms and
ranges of the $c\bar c$-$N$ potential. Our results show that nuclear
bound states of charmonium could be formed for $A \ge 6$. A value of
about 10~MeV is found for the binding of charmonium in nuclear matter,
which is comparable with the binding energy of nucleons, a remarkable
result since the nucleon-nucleon interaction is two orders of magnitude
stronger than the $c\bar c$-$N$ interaction. This is a consequence of
the absence of Pauli blocking in the charmonium-nucleon system.
Measurement of the $J/\psi$-nucleon scattering near threshold is
important to determine the van der Waals strength since this reaction is
dominated by gluon exchange \cite{BSdT}. Study of charmonium production
close to threshold \cite{Hoyer} and its interaction with nuclear
matter~\cite{Kharzeev} would provide a better understanding of the
mechanisms of color dynamics.

\begin{figure}
\caption
{\protect\label{ann}
$A_{NN}$ as a function of $p_{\rm lab}$ at $\theta_{\rm c.m.} = 90^\circ$ 
near the charm threshold, where the model is applicable. The data is taken from Ref.~4. The value of PQCD 
alone is 1/3. The upper, middle and lower curves correspond to 
$r_{\rm e} = 0.6 \rm\ fm$, $r_{\rm e} = 0.8 \rm\ fm$, and 
$r_{\rm e} = 1.2 \rm\ fm$ respectively and $a_{s} = -0.2 \rm\ fm$.
}
\end{figure}
\begin{table}
\caption
{\protect\label{tabla}
Binding energies $\langle H\rangle$ of the $\eta_c$ to various nuclei for a 
variational calculation corresponding to realistic nuclear densities
\protect\cite{Uberall} and a Gaussian form for the
 $c\bar c$-$N$ interaction. The values of $\langle H\rangle$
 are in MeV, the charmonium-nuclei reduced masses, $M_{\text{red}}$, are in 
GeV and the range of the potential, $R$, is in fm.
}
\begin{tabular}{ccccc}
& $M_{\text{red}}$ 
& $\langle H \rangle_{R = 0.4}$ 
& $\langle H \rangle_{R = 0.8}$
& $\langle H \rangle_{R = 1.2}$\\
\tableline
      ${}^4\text{He}$   &  1.66 &  $>0$   &  $>0$   &  $> 0$  \\
      ${}^6\text{Li}$   &  1.95 & $-0.12$ & $-0.07$ &  $> 0$  \\
      ${}^9\text{Be}$   &  2.21 & $-1.31$ & $-1.04$ & $-0.68$ \\
      ${}^{12}\text{C}$ &  2.36 & $-2.52$ & $-2.02$ & $-1.75$ \\
      ${}^{14}\text{N}$ &  2.44 & $-3.31$ & $-2.92$ & $-2.54$ \\
      ${}^{40}\text{Ca}$&  2.77 & $-6.13$ & $-5.83$ & $-5.31$ \\
      ${}^{56}\text{Fe}$&  2.83 & $-6.70$ & $-6.48$ & $-6.23$ \\
      ${}^{208}\text{Pb}$&  2.95 & $-9.24$ & $-9.20$ & $-9.12$ \\
\end{tabular}
\end{table}

\begin{references}
\bibitem{pie}
The value of $a_s = -0.2 \rm\ fm$ is 5 times larger than the value of 
Ref.~\cite{kaidalov} and \cite{Kharzeev} and 25 times that of 
Ref.~\cite{shevchenko}.
%
\smallskip
\end{references}
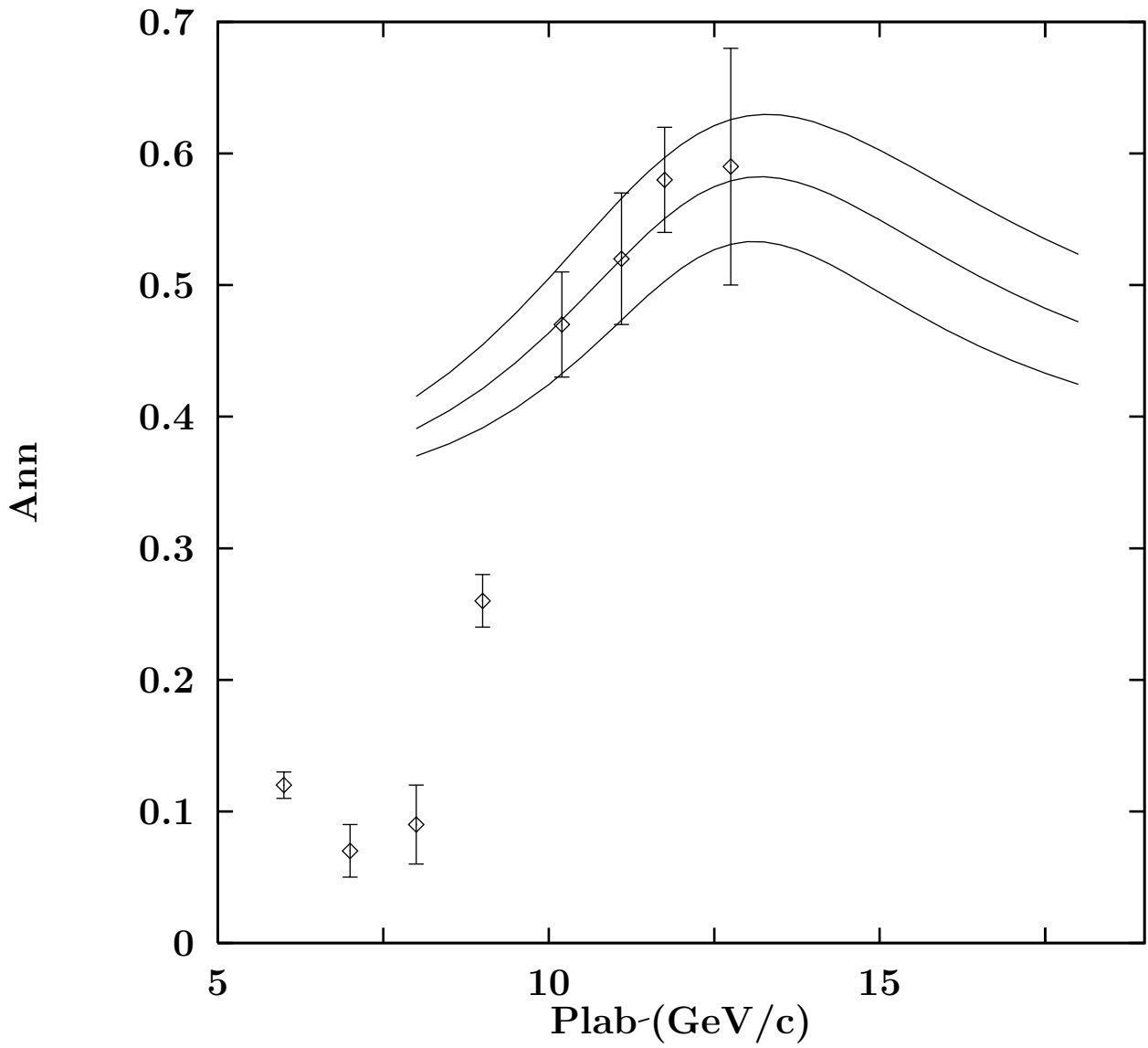
\end{document}